\documentclass[twocolumn,amsmath,amssymb,english,superscriptaddress]{revtex4-1}
\usepackage{graphicx,natbib,subfigure,tabularx,epsfig,longtable,amsfonts,rotating}
\usepackage{color,babel,currfile,subfigure,hyperref}
\begin{document}
\title{Machine Learning Phase Transition: An Iterative Proposal}
\author{X. L. Zhao and  L. B. Fu}\thanks{ E-mail: lbfu@gscaep.ac.cn}
\affiliation{Graduate School of China Academy of Engineering Physics, China}
\date{\today}

\begin{abstract}
We propose an iterative proposal to estimate critical points for statistical models based on 
configurations by combing machine-learning tools. Firstly, phase scenarios and preliminary 
boundaries of phases are obtained by dimensionality-reduction techniques. Besides, this step 
not only provides labelled samples for the subsequent step but also is necessary for its 
application to novel statistical models. Secondly, making use of these samples as training 
set, neural networks are employed to assign labels to those samples between the phase 
boundaries in an iterative manner. Newly labelled samples would be put in the training set 
used in subsequent training and the phase boundaries would be updated as well. The average 
of the phase boundaries is expected to converge to the critical temperature in this proposal. 
In concrete examples, we implement this proposal to estimate the critical temperatures for 
two q-state Potts models with continuous and first order phase transitions. Linear and manifold 
dimensionality-reduction techniques are employed in the first step. Both a convolutional neural 
network and a bidirectional recurrent neural network with long short-term memory units 
perform well for two Potts models in the second step. The convergent behaviors of the 
estimations reflect the types of phase transitions. And the results indicate that our proposal 
may be used to explore phase transitions for new general statistical models.
\end{abstract}
\maketitle
\section{Introduction}
\label{Introduction}
Machine learning is an advancement in computer science and technology adept at processing 
big data intelligently and has been applied to various applications in life such as image recognition, 
natural language processing and so on~\cite{Science349255,Nature521436,HBTNN3361,
Goodfellowbook}. In the field of theoretical condensed matter physics, the exponentially large 
Hilbert space usually makes calculations formidable for conventional approaches. With the 
advancing of computation power and the amelioration of algorithms, machine learning provides 
novel avenues to distill critical characteristics for configurations of statistical models and shows 
its talents in the context of classifying phases for condensed matter~\cite{NP13431,NP13435,
PRX7031038,PRL118216401,PRB96245119,PRL120066401,PRB94195105,PRB96195138,
PRE95062122,PRE96022140,PRB96144432}.  

In machine learning, architectures usually deal with massive samples which can be represented 
by points in feature spaces. Statistical characteristics can be analyzed by studying the distribution 
scenarios or patterns of such points, merely efficiency of analysis may be discounted due to 
redundant information and noise. Nevertheless, dealing with essential characteristic patterns 
would be conducive to improve the efficiency. The linear and manifold dimensionality-reduction 
techniques are frequently adopted in unsupervised machine-learning tasks to decrease redundant 
information and compress data~\cite{PM2559,IJPCA,NN11}. Due to their ability to reveal 
characteristic patterns for data sets, such techniques have been employed to investigate phase 
transitions and order parameters from the configurations of the classical Ising model
\cite{PRB94195105,PRE95062122, PRB96144432,PRB96195138,PRE96022140}. Such tools 
are suitable for representing data with essential information maintained as more as possible in 
a space with lower dimensionality without using supervised labels.

Different to unsupervised treatment, neural networks trained by samples with supervised labels 
are adept at classification tasks. During training, the parameters of a neural network would be 
updated in order to minimize a cost or loss function constructed by the output and supervised 
label. Such supervised learning tools have been employed to investigate Ising model~\cite{NP13431,
NP13435}, study strongly correlated fermions~\cite{PRX7031038,PRB94245129}, and explore 
the fermion-sign problem~\cite{SR78823}. Nontrivial topological states can also be learned by 
artificial neural networks. For example, by introducing quantum loop topography, a fully connected 
neural network can be trained to distinguish the Chern insulator and the fractional Chern insulator 
from trivial insulators~\cite{PRL118216401}. The phase boundary between topological and trivial 
phases can also be identified by a feed-forward neural network~\cite{PRB96245119}. Neural 
network can also be trained to learn the discrete version of the winding-number formula from a set 
of Hamiltonians of topological band insulators~\cite{PRL120066401}. The map between variational 
renormalization group and restricted Boltzmann machines has also been studied~\cite{arxiv14103831}. 
Quantum inspired tensor networks have been applied to multi-class supervised learning tasks and 
demonstrated by using matrix product states to parameterize models~\cite{ANIP294799}. An 
artificial neural network trained on entanglement spectra of individual states of a many-body 
quantum system can be used to determine the transition between a many-body localized and a 
thermalized region~\cite{PRB95245134}. The classification power of shallow fully connected and 
convolutional neural networks are examined by Ising model, concentrating on the effect of extended 
domain wall configurations~\cite{PRB97174435}. The generalizing and interpretability properties of 
machine-learning algorithms have been demonstrated by inferring the phase diagram for a liquid-gas 
transition~\cite{arxiv180702468}. These works indicate the power of neural networks in extracting 
the characteristics of the physical configurations.

As the works mentioned above, unsupervised learning techniques can be applied to cluster
physical objects in terms of the distributions in feature space. Supervised learning are employed 
based on the extracting ability of neural networks for physical samples. In this work, we propose 
an iterative proposal combing unsupervised learning techniques and supervised machine-learning 
architectures to estimate critical temperatures for statistical models only using their configurations 
with temperatures.  Firstly, dimensionality-reduction techniques are employed as unsupervised 
learning tools to obtain phase scenarios versus temperature by using the configurations (samples). 
Since no knowledge about phase transition is used, this is a necessary step for estimating critical 
points. It provides the scenario of the configurations in a range of temperature and preliminary 
phase boundaries. Meanwhile, initial labels for samples are obtained which would be used to train 
the neural networks subsequently. This step is also necessary for its application for new statistical 
models. Secondly, training the neural networks until they learn the characters of the existed samples, 
we use it to distinguish the samples with temperatures between the phase boundaries. Once the 
samples were recognized by the trained networks with high accuracy, they would be assigned the 
corresponding label and added into the training set to be used in subsequent training. Meanwhile, 
the phase boundaries would be updated by the temperatures of the last labelled samples. Without
extra knowledge about scaling behavior is used, we set the average value of the phase boundaries 
as the estimation for critical points. Iterating the procedure of recognizing and updating above, the 
estimation for critical points is expected to converge to the critical temperatures. Q-state Potts 
models provide a testing ground for various methods in the study of critical point theory
\cite{RMP54235,PCPS48106}. We demonstrate this proposal by two q-state Potts models with 
continuous and first order phase transitions, respectively. 

Following in Sec.~\ref{IMY}, we describe the iterative proposal of sequentially combing
dimensionality-reduction techniques and neural networks to estimate critical points. In Sec.
\ref{ImplementMds}, we employ this proposal to estimate the critical points for two q-state Potts 
models with continuous and first order phase transitions and analyze the results. Finally, we 
summarize in Sec.~\ref{Secconclusion}. 
\section{The Iterative Proposal}
\label{IMY}
There are two sequential steps in this proposal. The first one plays a necessary role in providing 
preliminary phase boundaries and supervised labels used in the second step. The details are listed 
in the following.
\subsection{Step-1: Dimensionality-reduction pretreatment}
\label{ssPCA}
This proposal starts from raw configurations of statistical models without extra knowledge about phase 
transitions. The configurations can be represented by points distributed in a high dimensionality feature 
space with cluster behaviors related to phases. It is a necessary step to obtain preliminary phase 
scenarios and phase boundaries in this proposal. However high dimensionality usually makes it 
burdensome to analyze statistical models. Thus it is desirable to employ a simpler or more accessible 
representation which maintains essential information of the distributions as much as possible. 
Dimensionality-reduction techniques offer such tools to extract salient characteristics for distributions
\cite{PM2559,IJPCA,NN11}. Scenarios of phases can be reflected by the distribution behaviors of 
the projected points in a lower dimensionality space. The phases can be used as supervised labels 
for the configurations in supervised machine learning. The initial phase boundaries which confine the 
critical temperature are identified preliminarily. This pretreatment not only provides essential 
cornerstone for the subsequent step but also is indispensable for its application for unexplored 
statistical models when only raw configurations with temperatures are provided. 
 
In this work we would employ singular value decomposition and multidimensional scaling which can 
be used in linear and manifold dimensionality-reduction tasks to do the unsupervised pretreament 
in the first step. However the choices of unsupervised techniques in this step are not limited to 
particular ones but any one which can be used to handle unsupervised clustering tasks may 
be a candidate. For example, uniform manifold approximation and projection which is based in 
Riemannian geometry and algebraic topology can also act as the unsupervised module here
\cite{arxiv180702468,arxiv180203426}; Isometric mapping algorithm represents the original data set 
on a quasi-isometric, low-dimensional space by preserving the geodesic distances between points on 
a neighborhood graph. This algorithm can also play the role above~\cite{Science2902319}; Besides, 
locally linear embedding which is used in manifold learning to obtain low-dimensional, 
neighborhood-preserved projections for high dimensional data sets can also be a candidate in this 
step~\cite{LLE}. These alternatives give this proposal generalization to be applied to new statistical 
models in term of distributions.

\begin{figure*}[t]
  \begin{center}
    \includegraphics[scale=0.45]{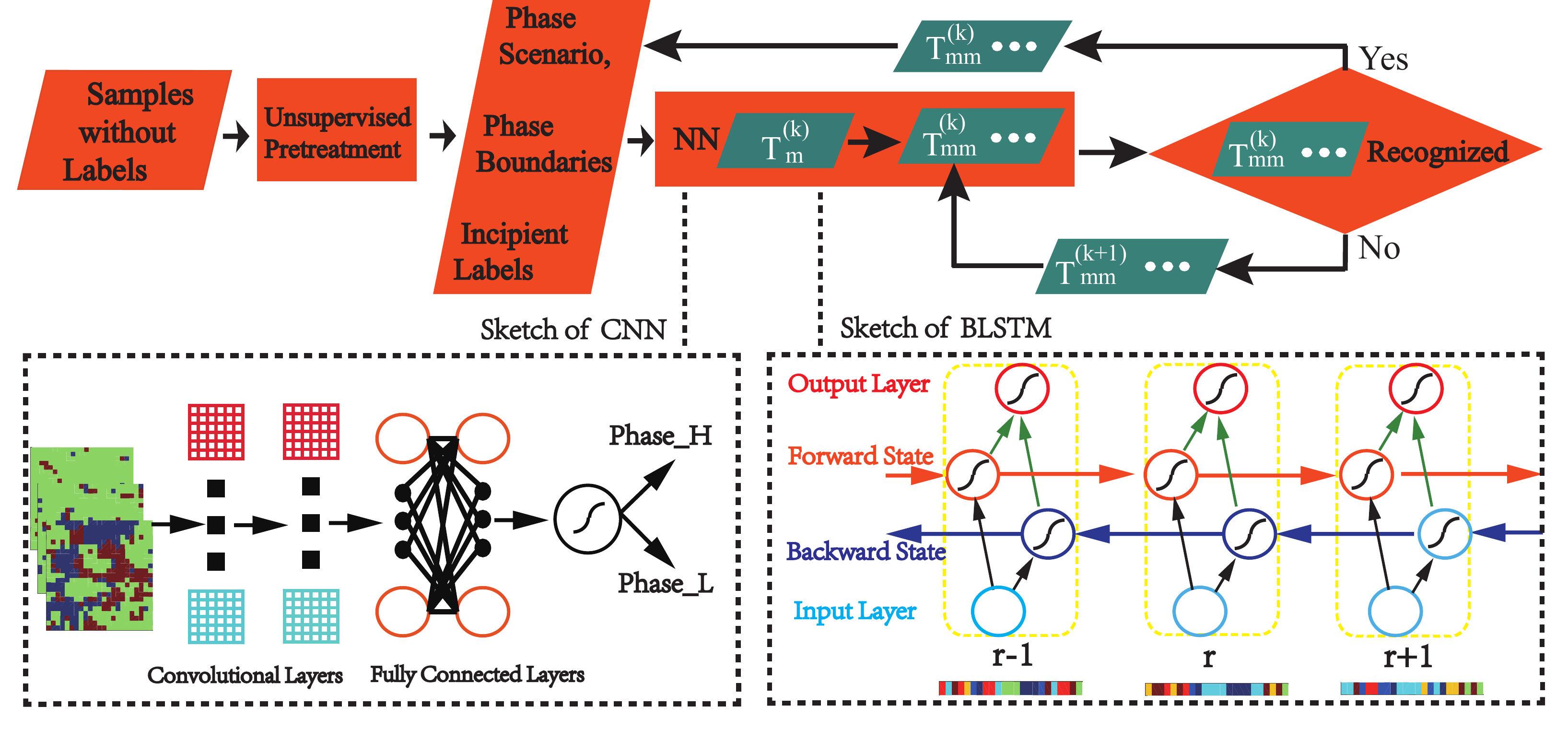}
    \vspace{0cm}
    \caption{The sketch of this proposal to estimate critical temperatures. The critical structure of the 
    convolutional neural network (CNN) and the bidirectional recurrent neural network with long short-term 
    memory units (BLSTM) are shown. The CNN deals with the batches of configurations. Whereas 
    configurations within the batches are split into rows as input denoted by $r$ when the BLSTM is 
    employed.} 
 \label{fig:Sketch}   
  \end{center}
\end{figure*}

Due to thermal fluctuations and information loss in the dimensionality-reduction treatment, those 
configurations between the phase boundaries would not be easily distinguished in the vicinity of critical 
points in this step. Benefiting from the pattern-extraction capability of neural networks, in the subsequent 
step~\ref{LocatingPP}, we give the procedure employing such supervised learning architectures to 
estimate critical temperatures for statistical models further based on the results of Step-1.

\subsection{Step-2: Estimating critical temperature iteratively}
\label{LocatingPP}
According to the results of Step-1, the critical point locates within the temperature interval where 
the labels of those samples have not been determined. In the following, we employ neural networks 
usually used in supervised learning tasks to estimate the critical temperatures for statistical models 
more accurately in an iterative way. Although we consider the situation that two phases are distinguished 
in Step-1, phase scenarios with more than two phases can be analyzed in segments with two phases. 
The workflow of this proposal is illustrated in Fig.~\ref{fig:Sketch}. The details for the second step are 
listed below.

$Step_1$: $T_{low~boundary}$, $T_{high~boundary}$ and $T_m=(T_{low~boundary}+T_{high~boundary})/2$
are used to denote the temporal low-temperature-phase boundary, high-temperature-phase boundary, 
and the average for the boundaries (estimation for critical temperature $T_c$), respectively. Taking the 
samples labelled in the first step~\ref{ssPCA} as preliminary training set, a neural network is trained until 
it can distinguish the whole training set with accuracy larger than a high threshold $S_{1t}$ within the 
number of $N_{1t}$ training batches used in training;

$Step_2$: If $Step_1$ is fulfilled, we use the trained neural network to recognize the samples at 
temperature $T_m$. If they are judged to be in \lq{}low-temperature phase (high-temperature phase)\rq{} 
according to a criterion $\mathcal{C}_m$, we further judge those at temperature 
$T_{mm}=(T_{low~boundary}+T_{m})/2$ (or $T_{mm}=(T_{high~boundary}+T_{m})/2$) according to a 
criterion $\mathcal{C}_{mm}$ by the same trained neural network; Else, if $Step_1$ is not fulfilled, we 
purge the newly added samples from the training set in the last iteration and roll $T_{mm}$ back to the 
last value to continue the recognition process from $Step_1$;

$Step_3$: If samples at $T_m$ are judged to be in \lq{}low-temperature phase (high-temperature phase)\rq{} 
and those at temperature $T_{mm}$ are also judged to be in \lq{}low-temperature phase (high-temperature 
phase)\rq{}, we assign the label to the samples at $T_{mm}$ as \lq{}low-temperature-phase label 
(high-temperature-phase label)\rq{} and add the newly labelled samples into the training set, to be used in 
following iterations. Meanwhile we update $T_{low~boundary}=T_{mm}$ (or $T_{high~boundary}=T_{mm}$); 
Else, we update $T_{mm}=(T_{low~boundary}+T_{mm})/2$ (or $T_{mm}=(T_{high~boundary}+T_{mm})/2$) 
and carry out the procedure in this step until the samples at $T_{mm}$ are assigned a label with the criterion 
$\mathcal{C}_{mm}$. Then we turn to $Step_1$ to begin the next recognition iterations.

Indeed, this proposal gives an iterative way to assign labels with high accuracy from margins towards center. 
When phase boundaries approach $T_c$, thermal fluctuations make the characteristics contained in the 
samples become harder to be learned by the neural networks with high accuracy. Even so, the newly added 
samples with labels enrich the training set and benefit the neural networks learning characteristics. In $Step_1$, 
the threshold of recognition $S_{1t}$ is necessary to be set large and $N_{1t}$ is used to avoid overfitting. 
The criteria $\mathcal{C}_{m}$ and $\mathcal{C}_{mm}$ have bearing on assigning labels and we give the 
concrete forms in the concrete examples following. 

Compared to the dimensionality-reduction pretreatment in the first step~\ref{ssPCA}, the function of the 
neural networks is more critical in terms of the precision of estimation. However the neural networks work 
based on the results of Step-1.

\section{Application to q-state Potts models}
\label{ImplementMds}
Now we turn to implement the proposal above to estimate the critical temperatures for two q-state Potts 
models on squire lattice with continuous and first order phase transitions respectively~\cite{RMP54235,PCPS48106}.

\subsection{Q-state Potts models}
\label{qPottsM}
Q-state Potts models are generalizations of Ising model with rich contents and offer agents to study 
ferromagnet and certain other physics of solid states~\cite{RMP54235,PCPS48106} . The general 
Hamiltonian for such models reads:
\begin{eqnarray}
\begin{aligned}
H = -J \sum_{\{i,j\}} \delta_{\sigma_{i},\sigma_{j}},
\end{aligned}\label{eq:Ising}
\end{eqnarray}
where the coupling $J$ between nearest-neighbor spins is set as the energy unit in this work. 
$ \delta_{\sigma_{i},\sigma_{j}}$ is the Kronecker delta function with regard to the variables 
$\sigma_{i}$ and $\sigma_{j}$ which take one of $q=(0,...,q-1)$ values with equal probabilities on each 
lattice site. A nearest-neighbor pair of same spins carries an unity energy and others are zero. While 
$q=2$, the Ising model is recovered in two dimensional case~\cite{ZP31253,PR65117}. These models 
posses self-duality points at temperature $T_c$ given by $e^{1/k_BT_c}=1+\sqrt{q}$~\cite{RMP54235}. 
In two-dimensional case, continuous phase transitions occur at $T_c=1/\log(1+\sqrt{q})$ when $q\le4$, 
and first order phase transitions occur at this temperature when $q>4$~\cite{JPC6L445,PRSLSA358535,
RMP54235} with Boltzmann constant $k_B=1$. In this work, only such spin configurations generated by 
Monte Carlo method would be used to estimate the critical temperatures~\cite{JASA44335,RPP60487,
MCBook}. These models offer benchmarks to check the performance of our proposal. 

\begin{figure*}[t]
  \begin{center}
   \includegraphics[scale=0.45]{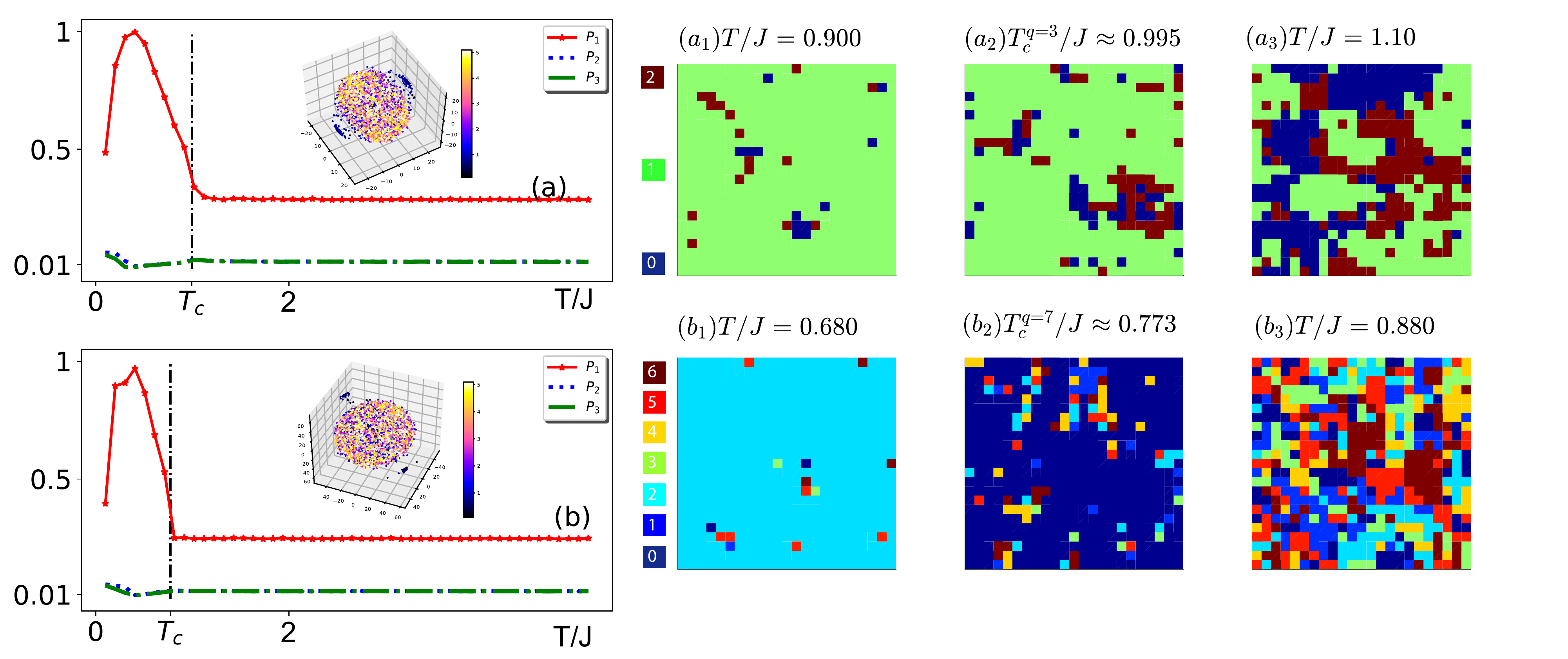}
    \vspace{-0.5cm}
    \caption{(a) the three leading normalized singular values ($P_1$, $P_2$ and $P_3$) as a function of 
    temperature for 3-state Potts model with the result of Metric-MDS in the inset as describe in section 
    \ref{SecSVD}. The leading singular values at each temperature are calculated by a vertical stack of 
    40 flattened configurations . (b) the results of SVD and Metric-MDS for 7-state Potts model. The 
    averages of the stacks have not been subtracted. This results to the bias in SVD but does not hamper 
    the usage of this technique to reveal the phase transitions. The temperatures of the samples are 
    mapped to the color of the projected dots. $(a_i)$ (i=1,2,3) list three configurations for 3-state Potts 
    model at different temperatures and $(b_i)$ (i=1,2,3) list those for 7-state Potts model. The codes 
    are given in the supplementary material~\cite{SpMl}. }
  \label{fig:TTT}
  \end{center}
\end{figure*}

Starting from the results of the first step~\ref{ssPCA}, we would employ a convolutional neural network 
(CNN) and a bidirectional recurrent neural network with long short-term memory units (BLSTM) to 
estimate the critical points for two q-state Potts models with different kinds of phase transitions\cite{Science349255,Nature521436,HBTNN3361,Goodfellowbook,PCPS48106,RMP54235}.

\subsection{Dimensionality-reduction techniques}
\subsubsection{Singular value decomposition}
\label{SecSVD}
According to section \ref{IMY}, firstly, we give a brief introduction to a technique used in linear 
dimensionality-reduction: singular value decomposition (SVD) and one of manifold learning techniques: 
multidimensional scaling (MDS). 

SVD is the core of principal component analysis to represent the original data in a lower 
dimensionality space where significant characteristics of statistical distributions are preserved
\cite{IJPCA,NN11}. While SVD reads $V=UZW^T$, the singular values are the diagonal elements of 
$Z$ usually normalized by dividing their summation and arranged in descending order ~\cite{IJPCA,NN11}. 
In this proposal $V$ is a vertical stack of flattened configurations in the form: 
$\mathbf{S}^{T}=(s_{11},...,s_{ij},...,s_{LL})^{T}$ with $s_{ij}$ denoting the spin value at site $(i,j)$ 
at temperature $T$. The original data are represented in a coordinate space where the covariance 
matrix is diagonalized by SVD. Since larger singular values together with the corresponding singular 
vectors carry more essential information of the original matrix, we will employ the three leading singular 
values to do the task for the first step~\ref{ssPCA}.

\subsubsection{Multidimensional scaling}
\label{SecMDS} 
No information is offered to indicate that linear dimensionality-reduction treatment can reflect the phase 
scenarios comprehensively here. For generality, we further employ MDS which is manifold learning 
technique to construct a lower dimensionality representation for the original data sets. There are mainly 
two kinds of MDS: one is called Non-Metric Multidimensional Scaling which preserves the ranks of the 
distances between the samples~\cite{SAGE,CMS,Borgbook,RDS133}; The other one is called Metric 
Multidimensional Scaling (Metric-MDS) which tries to reproduce the original Euclidean distances between 
pairs of points as well as possible in a lower dimensionality space and would be employed in this work. 
This method attempts to visualize the degree of similarities or dissimilarities for the investigated samples
\cite{SAGE,CMS,Borgbook,RDS133}.

Several loss functions can be constructed and optimized in Metric-MDS~\cite{SAGE,CMS,
Borgbook,RDS133}. For example, supposing a distance matrix ${\textstyle D^{\prime}=\{d_{ij}^{\prime}\}}$ 
with ${\textstyle d_{ij}^{\prime}=\sum_{k=1}^p{||x_{ki}^{\prime}-x_{kj}^{\prime}||}}$ in a reduced 
$p$-dimensionality space to approximate the Euclidean distance matrix ${\textstyle D=\{d_{ij}\}}$ of 
the pair-wise distances between samples $x_{i}$ and $x_{j}$ in the original space, a candidate loss 
function $Stress=\sum_{i,j}(\frac{||d_{ij}^{\prime}-d_{ij}||^2}{\sum_{i,j}d^{\prime2}_{ij}})^{1/2}$ can be 
minimized based on the differences between projected and original distances. Given the matrix 
${\textstyle D}$ for the original data set, a simple procedure for Metric-MDS follows:

$S_1$:~Appoint an arbitrary set of coordinates for projected points in a dimensionality reduced space with 
dimensionality $p$ ($p<4$, since distributions in spaces with dimensionality larger than 3 are difficult to be 
visualized);

 $S_2$:~Compute Euclidean distances among all pairs of projected points to form the matrix 
 ${\textstyle D^{\prime}=\{d_{ij}^{\prime}\}}$;

 $S_3$:~Evaluate loss function $Stress$ based on  ${\textstyle D}^{\prime}$ and ${\textstyle D}$;

 $S_4$:~Adjust coordinates $\{x_{k1}^{\prime},~x_{k2}^{\prime},~...,~x_{kN}^{\prime}\}$ to minimize 
 $Stress$;

 $S_5$:~Repeat the steps from $S_2$ to $S_4$ above until $Stress$ ceases decreasing.

$Stress$ is usually minimized by SMACOF (shorthand for Scaling by MAjorizing a COmplicated 
Function)~\cite{SAGE,CMS,Borgbook,RDS133}. And it decreases $Stress$ as a convex-optimization 
problem. Dimensionality-reduction treatment mentioned can be finished by Scikit-learn~\cite{Sklearn}

\subsection{Neural networks}
\label{NeuralNetW}
CNN and BLSTM are both artificial neural networks (ANN)~\cite{HBTNN3361,Goodfellowbook,
NN11,SSP1361}. In general classification tasks, the input $X_i$ (usually being a tensor fed into ANN 
and $i$ denotes classes) with the supervised label $Y_i$ would be used to train the ANN. $Y_i$ is 
usually treated as a distribution with $i$th component being one and others zero, in order to construct 
a cost function measuring the difference between the output of the ANN and the supervised labels. 
The parameters of ANN would be updated by algorithms to reduce the cost function. Convolutional 
operation plays an important role in CNN in company with nonlinear activation functions and some 
regulative operations. This operation extracts correlations in the samples in a region covered by 
convolutional filters with weights and biases to be updated. However in classification tasks, the rows 
(columns) in samples can be treated as ordered `elements' by recurrent neural networks. Thus the 
recurrent neural networks learn the samples in a global manner and long correlations can be extracted 
especially with LSTM units. In this proposal, once the samples with new temperatures between the 
phase boundaries are mapped into a certain category (phase) by the trained neural networks with 
high accuracy, they would be assigned the corresponding label and added into the training set. Thus 
along with the iterative process, the neural networks would be acquainted with the samples more and 
more accurately. To get an insight of the mapping process, an example of inputs, a batch of weights 
in the neural networks and the outputs are shown in the Appendix~\ref{appendix}. Different kinds of 
abstract characters would be extracted during training with the purpose to minimize the cost functions. 

\subsubsection{The convolutional neural network}
\label{ssCNN}
In terms of feature-map mechanisms of neural networks to extract characteristics of images, the 
configurations with similar patterns can be mapped into the same category with high accuracy\cite{Nature521436,HBTNN3361,Goodfellowbook,ANIPS1097}. Firstly, we elaborate the workflow 
of the CNN to learn the phases:
  Samples at $T/J$
  $\rightarrow$
  Convolutional operation-1
  $\rightarrow$
  ELU activation operation
  $\rightarrow$
  Max pooling
  $\rightarrow$
   Convolutional operation-2
   $\rightarrow$
   ELU activation operation
   $\rightarrow$
   Max pooling
   $\rightarrow$
   Fully connected operation-1
   $\rightarrow$
   ELU activation operation  
   $\rightarrow$
   Dropout operation
   $\rightarrow$
   Fully connected operation-2
   $\rightarrow$
   Calculate loss by processing cross-entropy cost function after Softmax operation
   $\rightarrow$
   Employ Adam algorithm to minimize the cost function.
   
Then we give a brief instruction to the operations in the CNN above as in Fig.\ref{fig:Sketch}. The critical 
operation is finished in convolutional layers through multiplying the pixels in a window of an input sample 
by the corresponding pixels in convolutional kernels cover the window, and biases are usually added. 
The kernels slide across the samples until all the pixels are visited to finish one time of convolutional 
map. The weights and the biases are shared during this operation. This reduces the amount of parameters. 
An example of convolutional operation is shown in the appendix~\ref{appendix}. To make the network 
be capable of extracting more kinds of features, several kernels are usually adopted. To extract more 
complex and abstract characteristics and improve the capability of recognition, deeper neural networks 
are recommendable candidates. 

After the convolutional operation, Exponential Linear Units (ELU): 
\begin{eqnarray}
ELU(x)=\left \{
\begin{array}{ll}
 x,~~~~~~~x>0, \\
e^{x}-1,~x<0, \\
\end{array}
\right.
 \label{eq:wind}
\end{eqnarray}
 would be employed as the activation function 
in this work~\cite{ELU}. Compared to the prevalent activation function Rectified Linear Unit (ReLU)
\cite{P27ICML807}, ELU has negative values which pushes the mean of activation closer to zero. This 
decreases the bias-shift effect and benefits machine learning with lower computational complexity
\cite{ELU}. Besides, the novel activation function: Swish performs better than ReLU in a number of 
scenes, and may be intriguing to be investigated in comparison with ReLU and ELU~\cite{arxiv171005941}. 
Max pooling operation converts high resolution matrix to lower resolution counterparts by outputting 
the maximum value in a sampling window. This operation reduces the amount of computation and 
preserves the essential information in a compact form. It not only makes feature detection more robust 
to scale and noise but also is beneficial to avoid overfitting, merely some information lost. Other 
versions of pooling, such as Average pooling or L2-norm pooling may also perform well
\cite{arxiv12070580}. 

Fully connected layers can be adopted to purify the information for next classification operations. 
Dropout is adopted not only to reduce overfitting but also to make the network less sensitive to small 
variations of inputs, benefiting the generalization of the network. This operation can be done by setting 
the values of some randomly chosen neurons in the network to zero~\cite{JMLR151929}. Softmax 
function: $Softmax(x)_i=\frac{e^{x_i}}{\sum_je^{x_j}}$ is employed to represent a probability distribution 
in this classification task. Finally, Adam algorithm (shorthand for Adaptive Moment Estimation) would be 
employed to minimize the cross-entropy cost function as a sum of loss functions: 
$Loss=-\sum_{q}y_{q}\log y_{q}^{\prime}$ in batches. Here $y_{q}$ and $y_{q}^{\prime}$ denote the 
supervised labels and the output (prediction) of a neural network, respectively~\cite{Kingma2014}. This 
is a computationally efficient algorithm with little memory requirements. It blends some advantages of 
Momentum algorithm, AdaGrad, and RMSProp algorithms~\cite{Goodfellowbook,Kingma2014}.

 \begin{figure*}[t]
  \begin{center}
    \includegraphics[scale=0.5]{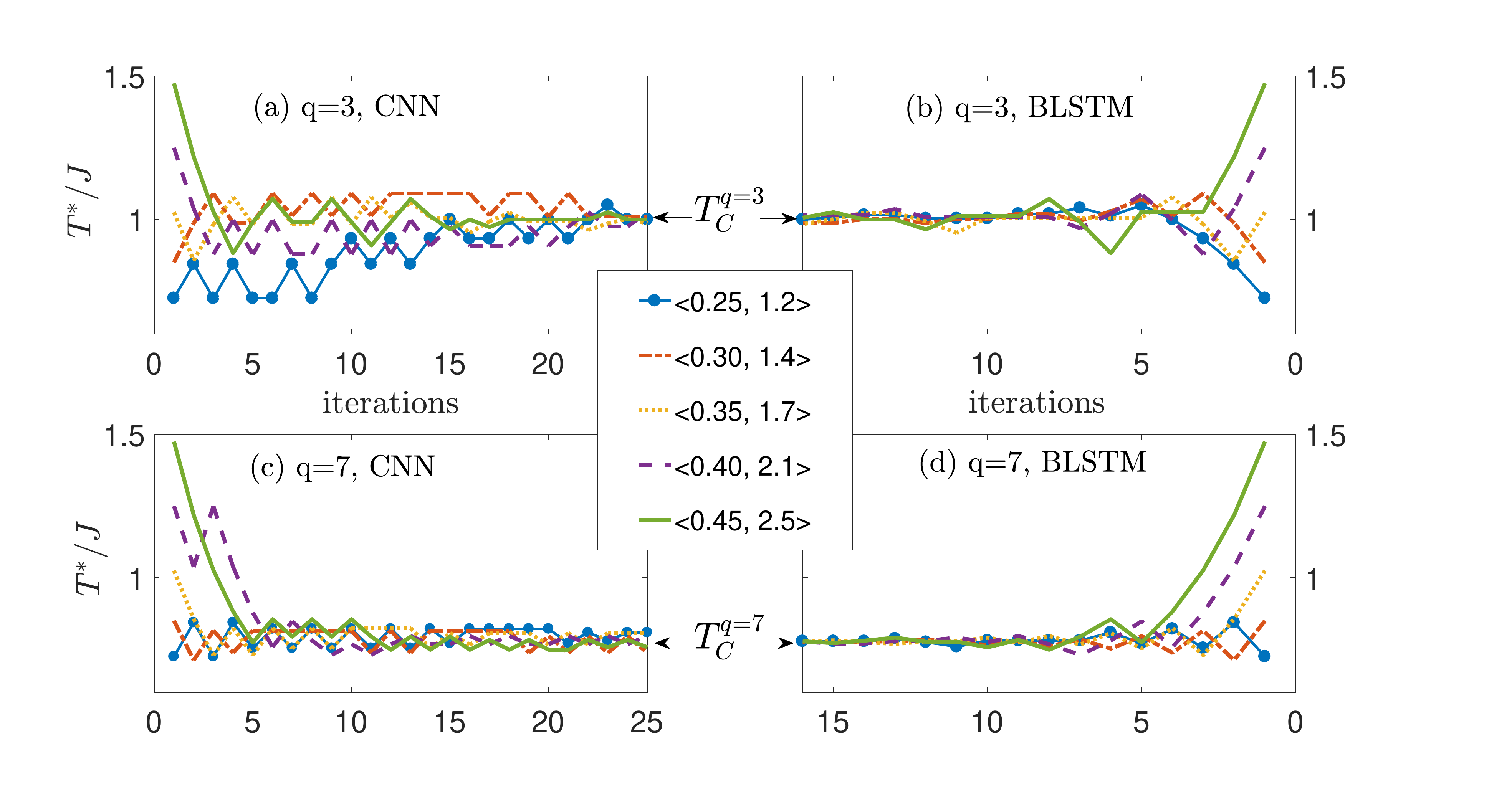}
    \vspace{-0.75cm}
    \caption{Trajectories of estimated critical temperatures with five different preliminary phase 
    boundaries versus the counts of the iterations from $Step_1$ to $Step_3$ for 3-state and 7-state 
    Potts models on $24\times 24$ lattices with CNN and BLSTM employed. 1200 preliminary samples 
    labelled in the the first step \ref{ssPCA} are evenly distributed between the temperature $T$ and 
    $T_{low(high)~boundary}$ when $(T_{low~boundary}-T)/J=0.2$ and $(T-T_{high~boundary})/J=0.2$, 
    respectively. In the second step \ref{LocatingPP}, $S_{1t}$=0.98 and $N_1t$=60 in $Step_1$. CNN: 
    $\mathcal{C}_{m}$ and $\mathcal{C}_{mm}$ are both that the labels are appointed to the output with 
    larger accuracy. 32 convolutional kernels of size $4\times4$ are used in the first convolutional layer 
    with strides=1 in both sliding directions. The initial weights and bias are set with a small amount of 
    random values. `SAME' padding is used in both the convolutional operations and max pooling. 64 
    convolutional kernels are adopted in the second convolutional layer. 1024 neurons are used in both 
    fully connected layers with initial weights and bias same to those of convolutional layers above. 0.5 
    probability of zeroing is set in the dropout operation. The output of the second fully connected layer 
    runs through the softmax operation, to construct the cross-entropy cost function which would be 
    minimized by Adma algorithm learning rate varies as $0.0001\times 0.99^{k}$ where $k$ denotes 
    the number of training batches cost. Max pooling takes $2\times2$ window, and strides 2;  BLSTM: 
    the criteria $\mathcal{C}_{m}$ and $\mathcal{C}_{mm}$ are that the labels are appointed to the 
    output with accuracy of 0.2 larger. Adma algorithm is also used to minimize the cross-entropy cost 
    function constructed by the output and labels with fixed learning rate 0.001 and 256 LSTM neurons 
    are used. The legends list the preliminary phase boundaries in the format:  
    $<T_{low~boundary},T_{high~boundary}>$ in units of $J$. TensorFlow has been used to implement 
    the neural networks~\cite{tensorflow2015}. The codes are given in the supplementary material
    \cite{SpMl}. } 
 \label{fig:Errorbar}   
  \end{center}
\end{figure*}

\subsubsection{The bidirectional recurrent neural network with long short-term memory units}
\label{ssRNN}
Recurrent neural networks have the framework of feeding the output of hidden layers back into 
themselves. This makes them expert in dealing with sequential data where the order of context is 
significant~\cite{NN6185}.

Like CNNs, RNNs can also be trained using Adam algorithms in back-propagation process. Merely, 
in trivial RNNs, gradient usually vanishes for long distance between input and target which limits 
their applications from learning long-term dependencies~\cite{arxiv12115063}. LSTM was designed 
to remedy this problem with memory blocks regulating the flow of information instead of simple 
sigmoid cells: $Sigmoid(x)=\frac{e^{x}}{(e^x+1)}$, in the hidden layers~\cite{JMLR3115,NC91735}. 
The blocks consist of cells with a forget gate, an input gate and an output gate to control the influence 
of past on current flow, input and output by multiplying the corresponding weights with biases usually 
added. Sigmoid functions are usually used to regulate the results in these gates~\cite{JMLR3115}.

Unlike trivial RNNs, bidirectional recurrent neural networks were invented to utilize the information 
both from the past and future of a sequence by merging two RNNs with information propagating in 
opposite directions into one output~\cite{IEEE4511,IEEE799}. This structure makes the networks 
learn more correlative information of training sequences and be most sensitive to the values around 
currently evaluate time point. During training, forward and backward states do not interact with each 
other. After forward and backward processes are done, the network would be updated by minimizing 
cost functions through back-propagation. Applying LSTM units in trivial bidirectional recurrent neural 
networks, one gains the BLSTM as shown in Fig.\ref{fig:Sketch} which we will use in this work.

To classify the configurations by a BLSTM, different to CNN, the order of rows of configurations 
acts as time step and each row is treated as the input at the corresponding time step. For the 
configurations with size $L_x\times L_y$, we handle $L_y$ sequences of length $L_x$ for each
configuration. One training iteration of the BLSTM follows:
Samples at $T/J$
$\rightarrow$
Forward and backward cells of LSTM merge into one output
$\rightarrow$
Calculate loss by processing cross-entropy cost function with Softmax operation
$\rightarrow$
Employ Adam algorithm to minimize the cost function.
   
The function of BLSTM in this proposal is same to the CNN above. Besides these two architectures, 
other powerful neural networks, such as variations of LSTM neural networks, ResNet and deeper 
neural networks are also highly recommended to act as such a module in this proposal to extract 
characteristics for general statistical models~\cite{HBTNN3361,Goodfellowbook,arXiv151203385}.

\subsection{The estimating results}
\label{estimatingresult}
As shown by the results of SVD in Fig.~\ref{fig:TTT} (a) and (b), the configurations for 3-state and 
7-state Potts models can be divided into two phases versus temperature which are confirmed by 
the results of Metric-MDS in the insets. It can be seen that the critical points locate near $T/J=1$ 
for 3-state Potts model and $T/J=0.8$ for 7-state Potts model which approximate theoretical value
$T_c=1/\log(1+\sqrt{q})$. This pretreatment also offers the thresholds for the preliminary phase 
boundaries. At high temperature phases, the disordered configurations with spin values which 
distribute obeying detailed balance, lead to the plateaus of $P_1$. This corresponds to the isotropic 
distributions of the projected points in the Metric-MDS. Both at very low temperature region, the 
neighbor spins tend to take same values and when temperature is very high, the neighbor spins 
tend to be different. In both the cases, the fluctuations distribute homogeneously across the lattice 
which results to the low value of $P_1$. Thus it is necessary to use Metric-MDS to reveal the 
phase scenarios more discreetly. At low temperature phase, the projected points tend to form 
centers along the central axes of the distributions in Metric-MDS. 

The obtuse and sharp turning behaviors of $P_1$ near the critical points correspond to the physical 
features of continuous and first order phase transitions of 3-state and 7-state Potts models, 
respectively. While there is no more than one dominant maximal singular value in the temperature 
region, the first two leading principal components are enough to reveal the phase scenarios if one 
desires to use principal components analysis in the first step. Merely the data should be centered 
by subtracting the mean values. 

According to the results above, the thresholds for the preliminary boundaries for low and high 
temperature phases can be set as $T_{boundary}$=0.25 and 1.2 in units of $J$, respectively. 
Then the samples with temperatures outside the region ($0<T/J<0.25$ and $T/J>1.2$) can be 
assigned low and high temperature labels, respectively. Even there is fluctuation to choose these 
boundaries, supervised learning tools would be employed to compensate this indeterminacy. Thus 
different initial phase boundaries would be checked to show the validity of this proposal. Starting 
from the above results, the critical temperatures for 3-state and 7-state Potts models can be 
estimated according to the iterations from $Step_1$ to $Step_3$ in~\ref{LocatingPP}. In 
Fig.\ref{fig:Errorbar}, along with the iterations, the estimations for the critical temperatures 
converge to the theoretical values in all the cases. With the same parameters, the critical 
temperature of 7-state Potts model is more easier to be estimated than that of 3-state Potts 
model both for the neural networks being CNN and BLSTM. This performance may be related to 
the characteristics of continuous and first order phase transitions. For the samples on two sides 
in the vicinity of the critical points, the characteristics are more legible to be learned for first order 
phase transitions than continuous phase transitions. This corresponds to the behaviors of $P_1$ 
near the critical temperatures in Fig.\ref{fig:TTT}. However we do not exclude the influence of 
$q$ values to the learning results of the neural networks. Although both the CNN and the BLSTM 
perform well at estimating the critical temperatures for both the models, the estimation by BLSTM 
converges more quickly than that by CNN. Merely there is little comparability between these 
results since their mechanisms of recognition are different. This hints that other supervised 
machine-learning tools with different learning mechanisms may also perform well. And the results 
indicate that it may be worthy to investigate the relation between the performance of machine-learning 
tools in learning statistical models and the universal scaling behaviors for those models.

\section{Summary}
\label{Secconclusion}
In this work, we show the proposal combining techniques used in dimensionality-reduction processes
and neural networks used in supervised learning to estimate critical points for statistical models. And 
the proposal is demonstrated by two q-state Potts models with different kinds of phase transitions. 
Actually, this proposal is an iterative approach to assign labels from margins towards center with high 
accuracy only starting from configurations with corresponding temperatures. The unsupervised 
pretreatment in the first step aims to offer preliminary phase boundaries and training samples with 
supervised labels which would be used subsequently by neural networks to estimate critical 
temperatures more accurately. The determination of phase scenarios and preliminary phase boundaries 
in the first step is not only necessary for the subsequently step but also indispensable for its application 
on similar problems. The training set would be enriched during the iteration process, which makes the 
neural networks be acquainted with the samples more and more accurately. In concrete examples, 
both a convolutional neural network and a bidirectional recurrent neural network with long short-term 
memory units perform well at estimating the critical points for two q-state Potts models. With the 
same neural network, the performances of the estimations for the models correspond to the features 
of the phase transitions which hints that machine-learning tools may be used to reflect universal scaling 
laws for statistical models by scientific approaches in the future. Since other machine-learning 
architectures can also be employed as modules in this proposal, we anticipate that this proposal may 
work to study new statistical models. 

\section*{Acknowledgements}
The Xiaolong Zhao thanks Hongchao Zhang at Dalian University of Technology for helpful discussions. 
This work is supported by the National Natural Science Foundation of China (Grant No. 11725417, 
11575027), NSAF (Grant No. U1730449), and Science Challenge Project (Grant No. TZ2018005).

\section*{APPENDIX}
\label{appendix}
Updating the parameters like weights of the neural networks play an important role in the learning 
process. The trained neural networks act like a kind of map architecture. We illustrate the mapping of 
the networks by a couple of inputs at different phases, weights of trained neural networks and the 
corresponding outputs. The inputs (matrix here), after multiplied by the matrices of weights in particular 
manner (like convolutional operation, bias added) up to some regulative activation operations (activation 
functions, pooling and so on as in section~\ref{NeuralNetW}), would be mapped to the corresponding 
categories (phases). The weights and biases can be updated by algorithms like Adaptive Momentum 
(adopted in this work),  Adaptive Gradient, Momentum and so on. The operations like convolutional 
operation and structures like gates in LSTM units, make the neural networks extracting characteristics 
effectively. 
 \begin{figure*}[t]
  \begin{center}
    \includegraphics[scale=1]{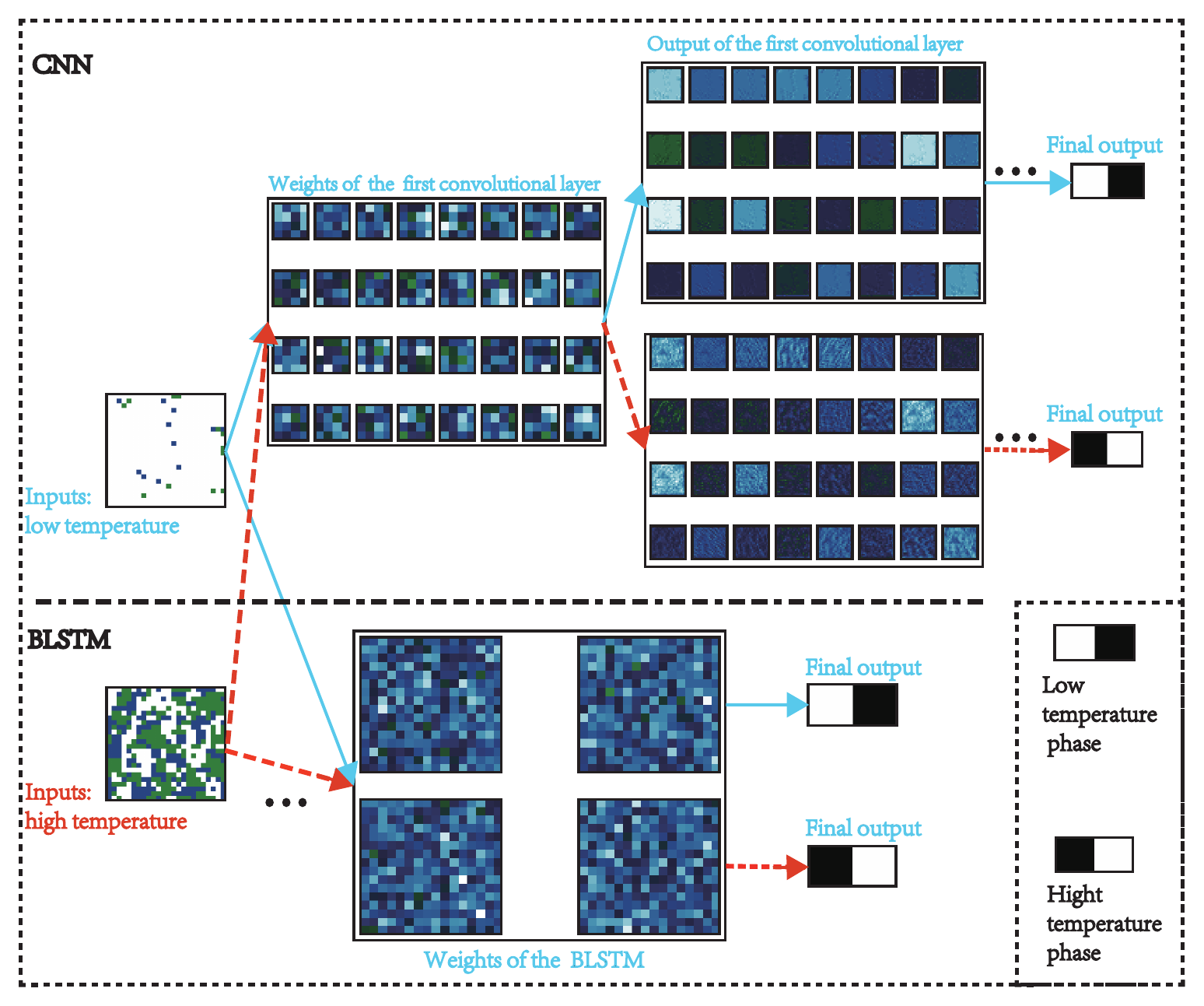}
    \vspace{0cm}
    \caption{The part above the dot-dashed line: a batch of 32 filters (constructed by weights) in 
    the first convolutional layer and the outputs for two samples at temperatures ($T/J=0.8$) and 
    ($T/J=1.5$) for q=3 Potts model after the CNN has been trained. The part below the dashed line 
    are the weights (reshaped) in the BLSTM. The networks are same to those in Fig.~\ref{fig:Errorbar}. 
    The codes are provided in the supplementary material~\cite{SpMl}. } 
 \label{fig:Appendix}   
  \end{center}
\end{figure*}

\end{document}